\documentclass[12pt,preprint]{aastex} 

\usepackage{natbib}
\usepackage{epsf}
\usepackage{url}
\usepackage{color}
\newcommand{\Msun}{\ifmmode\mbox{M}_{\odot}\else$\mbox{M}_{\odot}$\fi}
\newcommand{\Rsun}{\ifmmode\mbox{R}_{\odot}\else$\mbox{R}_{\odot}$\fi}
\newcommand{\Mearth}{\ifmmode\mbox{M}_{\oplus}\else$\mbox{M}_{\oplus}$\fi}
\newcommand{\Rearth}{\ifmmode\mbox{R}_{\oplus}\else$\mbox{R}_{\oplus}$\fi}

\shorttitle{UBC AI}
\shortauthors{Zhu et al.}

\begin{document}
\title{Searching for pulsars using image pattern recognition}

\author{W. W. Zhu\altaffilmark{1},
A. Berndsen\altaffilmark{1},
E. C. Madsen\altaffilmark{1},
M. Tan\altaffilmark{1},
I. H. Stairs\altaffilmark{1},
A. Brazier \altaffilmark{2},
P. Lazarus\altaffilmark{3},
R. Lynch\altaffilmark{4},
P. Scholz\altaffilmark{4},
K. Stovall\altaffilmark{5,21},
S. M. Ransom\altaffilmark{6},
S. Banaszak\altaffilmark{7},
C. M. Biwer\altaffilmark{7, 19},
S. Cohen\altaffilmark{5},
L. P. Dartez\altaffilmark{5},
J. Flanigan\altaffilmark{7},
G. Lunsford\altaffilmark{5},
J. G. Martinez\altaffilmark{5},
A. Mata\altaffilmark{5},
M. Rohr\altaffilmark{7},
A. Walker\altaffilmark{7},
B. Allen\altaffilmark{8,7,17},
N. D. R. Bhat\altaffilmark{9,10},
S. Bogdanov\altaffilmark{11},
F. Camilo\altaffilmark{11,13},
S. Chatterjee\altaffilmark{2},
J. M. Cordes\altaffilmark{2},
F. Crawford\altaffilmark{12},
J. S. Deneva \altaffilmark{20},
G. Desvignes\altaffilmark{3},
R. D. Ferdman\altaffilmark{4,14},
P. C. C. Freire\altaffilmark{3},
J. W. T. Hessels\altaffilmark{15,16},
F. A. Jenet\altaffilmark{5},
D. L. Kaplan\altaffilmark{7},
V. M. Kaspi\altaffilmark{4},
B. Knispel\altaffilmark{8,17},
K. J. Lee\altaffilmark{3},
J. van Leeuwen\altaffilmark{15,16},
A. G. Lyne\altaffilmark{14},
M. A. McLaughlin\altaffilmark{18},
X. Siemens\altaffilmark{7},
L. G. Spitler\altaffilmark{3},
A. Venkataraman\altaffilmark{13}
}
\altaffiltext{1}{\footnotesize Department of Physics and Astronomy,
6224 Agricultural Road, University of British Columbia, Vancouver, BC, V6T 1Z1, Canada;
zhuww@phas.ubc.ca, berndsen@phas.ubc.ca}
\altaffiltext{2}{\footnotesize Astronomy Department, Cornell University,
Ithaca, NY 14853, USA}
\altaffiltext{3}{\footnotesize Max-Planck-Institut f\"{u}r Radioastronomie,
Auf dem H\"{u}gel 69, D-53121 Bonn, Gemnay}
\altaffiltext{4}{\footnotesize Department of Physics, McGill University,
Montreal, QC H3A 2T8, Canada}
\altaffiltext{5}{\footnotesize Center for Advanced Radio Astronomy, University of Texas at Brownsville, Brownsville, TX 78520, USA}
\altaffiltext{6}{\footnotesize NRAO, Charlottesville, VA 22903, USA}
\altaffiltext{7}{\footnotesize Center for Gravitation, Cosmology and
Astrophysics. University of Wisconsin Milwaukee, Milwaukee, WI 53211 USA}
\altaffiltext{8}{Max-Planck-Institut f\"{u}r Gravitationsphysik, D-30176
Hanover, Germany}
\altaffiltext{9}{\footnotesize International Centre for Radio Astronomy
Research, Curtin University, Bentley, WA 6102, Australia}
\altaffiltext{10}{\footnotesize Centre for Astrophysics \& Supercomputing,
Swinburne University, Hawthorn, Victoria 3122, Australia}
\altaffiltext{11}{\footnotesize Columbia Astrophysics Laboratory, Columbia
University, New York, NY 10027, USA}
\altaffiltext{12}{\footnotesize Department of Physics and Astronomy, Franklin
and Marshall College, PO Box 3003, Lancaster, PA 17604-3003, USA}
\altaffiltext{13}{\footnotesize Arecibo Observatory, HC3 Box 53995, Arecibo,
PR 00612, USA}
\altaffiltext{14}{\footnotesize University of Manchester, Jodrell Bank
Observatory, Macclesfield, Cheshire SK11 9DL, UK}
\altaffiltext{15}{ASTRON, the Netherlands Institute for Radio Astronomy,
Postbus 2, NL-7900 AA, Dwingeloo, the Netherlands}
\altaffiltext{16}{Astronomical Institute `Anton Pannekoek,' University of
Amsterdam, Science Park 904, NL-1098 XH Amsterdam, the Netherlands}
\altaffiltext{17}{Leibniz Universit\"{a}t Hannover, D-30167 Hannover, Germany}
\altaffiltext{18}{Department of Physics, West Virginia University Morgantown, WV
26506, USA}
\altaffiltext{19}{Department of Physics, Syracuse University, NY 13244 USA}
\altaffiltext{20}{Naval Research Laboratory, 4555 Overlook Ave SW, Washington, DC 20375}
\altaffiltext{21}{Department of Physics and Astronomy, University of New
Mexico, Albuquerque, NM, 87131}



\newcommand{\acro}{{PICS\,}}
\begin{abstract}
In the modern era of big data, many fields of astronomy are generating
huge volumes, the analysis of which can sometimes be the limiting
factor in research.
Fortunately, powerful data-mining techniques have been developed
by computer scientists, ready to be applied to various fields.
In this paper, we present a novel artificial intelligence (AI) program that
identifies pulsars from recent surveys using image pattern recognition
 with deep neural nets---the \acro(Pulsar Image-based Classification
System) AI. 
The AI mimics human experts and distinguishes pulsars from noise and
interference by looking for patterns from candidate plots. 
Different from other pulsar selection programs which searched for expected
patterns, 
the PICS AI is taught the
salient features of different pulsars from a set of human-labeled candidates
through machine learning.
The training candidates are collected from the Pulsar Arecibo L-band Feed Array
Survey.
The information from each pulsar candidate is synthesized in four
diagnostic plots, which consist of image data with up to thousands of
pixels. 
The AI takes these data from each candidate as its input and uses
thousands of such candidates to train its $\sim$9000 neurons.
The deep neural networks in this AI system grant it superior ability
in recognizing various types of pulsars as well as their harmonic signals.
The trained AI's performance has been validated
with a large set of candidates from a different pulsar survey, the Green Bank
North Celestial Cap survey.
In this completely independent test, 
PICS ranked 264 out of 277 pulsar-related candidates, including all 56 previously known
pulsars and 208 of their harmonics, in the top 961 (1\%) of 90008 test candidates, missing only 13 harmonics.
The first non-pulsar candidate appears at rank 187, following 45 pulsars and 141 harmonics.
In other words, 100\% of the pulsars were ranked in the top 1\% of all
candidates, while 80\% were ranked higher than any noise or interference.
The performance of this system can
be improved over time as more training data are accumulated. This AI system has been integrated into the PALFA survey
pipeline and has discovered six new pulsars to date.
\end{abstract}

\section{Introduction}
\label{sec:intr}

Recent pulsar surveys such as the Pulsar Arecibo L-band Feed Array (PALFA;
\citealt{cfl+06,kas12,laz12}) survey and the Green Bank North Celestial Cap
(GBNCC; \citealt{2013IAUS..291...41L},  K. Stovall et~al. in prep.)
survey are expected to find---or are already finding---hundreds of new pulsars among many
millions of pulsar candidates. However, the surveys are polluted by radio frequency
interference (RFI) that makes it hard to pick out the pulsars from the
candidates produced using simple
metrics such as the signal-to-noise ratio (SNR). Human experts can look at diagnostic
plots of the candidates and identify the pulsars more successfully, but it is impractical 
to inspect millions of candidates that way. In this paper,
we present an artificial intelligence (AI) system that emulates human
experts and classifies pulsar candidates using patterns from four standard
diagnostic plots---the pulse profile, time-versus-phase plot,
frequency-versus-phase plot, and dispersion-measure (DM) curve (see Section~\ref{sec:features} for details). This system was
trained with PALFA candidates classified by human experts, its
parameters were tuned using a cross-validation set of candidates, and its
final performance was compared against a large
set of manually identified candidates from the GBNCC survey (project code:
GBT09C-057).
Both the PALFA and GBNCC candidates are generated using a pipeline that
runs the {\it PRESTO}\footnote{http://www.cv.nrao.edu/$\sim$sransom/presto/}
search software \citep{ran01,rem02},
but other pulsar-searching pipelines, such as the one in 
the Einstein@Home project \citep{akc+13,kek+13}, can produce these same diagnostic plots, so it is possible to apply our AI system to most pulsar surveys with little modification.

In the past, several successful candidate-sifting schemes have been developed
for different surveys. Some involve graphical interfaces which
allow for the interactive selection of pulsar candidates based on the pulse period and
SNR~\citep{fsk+04}. Some apply heuristic
scoring algorithms to the candidate diagnostic plots, using
statistical tests, curve fitting, and a graphical interface to visually
inspect the distribution of pulsar scores in the scores' parameter space
\citep{kel+09}. 
An effective sorting scheme was constructed \citep{Lee09, lsj+13} using a combination of six carefully designed heuristic scores. One particular score compares the candidate's pulse frequency against the frequency distribution drawn from a large sample of the survey
candidates; recognizing that the majority of candidates are RFI, this single
histogram removes a large fraction of the repeatedly observed RFI,
especially harmonics of the 60~Hz interference from the power supply. 
\citet{emk+10} improved the method of \citet{kel+09} by
applying machine learning (ML) to the heuristic scores. Instead of inspecting the score distribution by eye, they fed the scores into an
artificial neural network and trained the network to classify candidates.
\citet{bbb+12} expanded the number of scores used by fitting the
candidate's features with different model curves, and also used a neural
network to combine these scores. 
Another comprehensive score-based system \citep{kas12, laz12} was developed
and has been used to find many pulsars for the PALFA survey. 
Most recently, \citet{kek+13} designed algorithms that check for outstanding
signals in some of the diagnostic plots by binning the plots with pre-designed
patterns in the shape of vertical lines or area patches, and applied them in the Einstein@Home project. 

Notably, these previous candidate-sifting systems use heuristically-designed
functions which characterize patterns in the diagnostic plots into a set of
scores. Such score-based systems have some advantages. 
They make good use of
 the candidate's properties like period, DM and computed information like
the significance of the periodicity and the width/height of the summed profile. 
However, such systems often rely on matching candidates with
some pre-designed patterns, such as a Gaussian-like peak.
Some of these designed patterns 
do not match pulsars with multiple pulse peaks well.
When these designed patterns do match the candidate, they tend to average out
the details and small features in the diagnostic plot.
These small features can sometimes be very useful in distinguishing pulsars
from RFI.
Some score-based
systems select pulsar candidates by drawing the score distributions from known
pulsars. Such systems may be biased against rare types of pulsars, such as pulsars
with wider profiles, skewed DM curves or low signal strength. 

\begin{figure}
\includegraphics[scale=0.4]{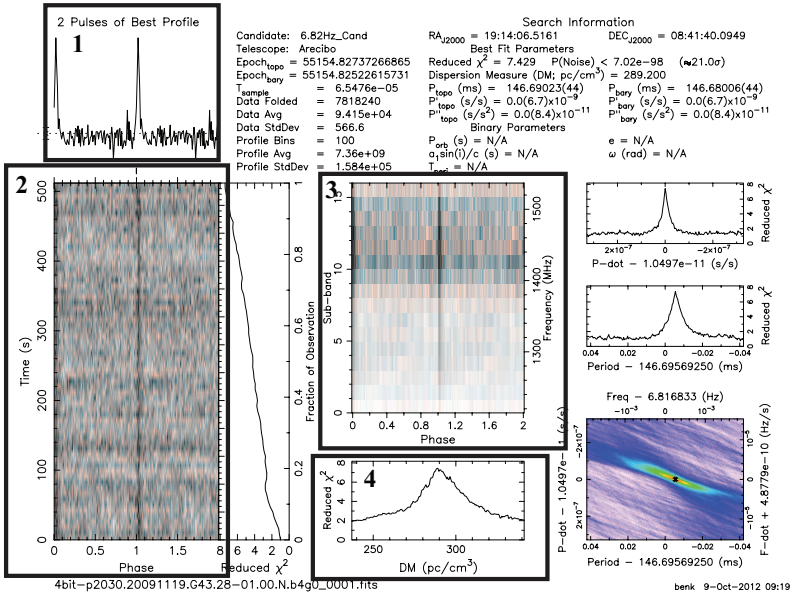} \\ 
\caption {\label{fig:presto} 
{\it prepfold} diagnostic plot for PICS-discovered pulsar PSR
J1914+08. The four key subplots of the PICS AI system are highlighted: 1. summed pulse profile 2. time-versus-phase plot 3. frequency-versus-phase plot 4. DM curve. For subplot 1, 2 and 3
the pulse phase are wrapped around twice to show two duplicated pulses.
}
\end{figure}

In contrast, the \acro AI applies image pattern recognition directly to the
original diagnostic plots, and determines what patterns to match through
machine learning.
Using the original diagnostic plots allows the AI to utilize the detailed
information in the plots. Using
ML allows us to train the AI with a wide variety of pulsar
candidates. A significant fraction of our training candidates are pulsars or
their harmonics with weak, broad, or multi-peak pulse profiles (for example, a
harmonic signal at $1/2$ period of the fundamental would be a factor of $\sim
2$ broader and $1/\sqrt{2}$ weaker and a harmonic at $2\times$ period would have two
peaks). 
As a result, the \acro AI is sensitive not only to pulsar candidates but also to their harmonics.
In fact, one of the new pulsars discovered by PICS, PSR J1914+08 (Figure~\ref{fig:presto}), was identified from its $6/11$, $6/7$ and $2/3$ harmonics. In this case, the fundamental frequency was missed by the PALFA pipeline due to RFI.
Finally, an important feature of the \acro AI is that it neglects information such as the candidate's period, DM and sigma value.
Instead, it focuses on only the details in the normalized diagnostic plots.  
This feature makes it a good complement to the score-based systems.

This paper introduces a new approach to the candidate-sifting problem using
supervised ML based on image patterns. In Section~\ref{sec:imp},
we describe the candidate plots which contain the physical features
distinguishing them as pulsars, and we detail the data-preparation process
required to adapt these plots as inputs to the ML system. We also
describe the structure of the \acro AI system. In
Section~\ref{sec:res}, we detail the AI's test performance and results from
classifying GBNCC data. Finally in Section~\ref{sec:dis}, we discuss the
results and the AI system's unique strengths and features.

\section{Implementation}
\label{sec:imp}

The diagnostic plot that a human expert relies on to identify a pulsar contains
several important subplots. 
In order to combine the information from these subplots, we constructed the PICS AI with a two-layer hierarchy (Figure \ref{fig:layers}). 
The first layer consists of a group of ML classifiers trained to
look at different subplots, providing a pool of experts capable of
recognizing pulsar patterns; while the second layer combines the
classifications from first layer to classify the candidate.
Each first-layer classifier rates how ``pulsar-like'' a particular subplot of the
candidate is with a number between 0 (not a pulsar) and 1 (a pulsar), giving a
prediction matrix that is the output of the first layer. These votes are fed into a second-layer classifier which learns to
properly weight these votes and forms a final consensus on how
pulsar-like a candidate is. 

In this section we provide a detailed discussion of the PICS AI implementation\footnote{The AI source code and a trained classifier are accessible on
github\url{https://github.com/zhuww/ubc_AI}}, starting from the most important subplots that the AI uses.

\subsection{Four features}
\label{sec:features}

Our goal is to train an AI program that mimics human
experts. Here we first introduce how the PALFA pipeline finds pulsar candidates
and then discuss which features human experts look for when identifying promising candidates.

The PALFA survey uses the 7-beam L-band (1.4 GHz) receiver at the Arecibo
observatory. The survey takes 5-minute snapshots of the Galactic plane. In
recent observations, data were taken using the Mock spectrometer, which has
a 65.5 $\mu$s sampling time and 960 frequency channels covering 322.6 MHz
of bandwidth. The data from each beam are
analyzed using a {\it PRESTO}-based pulsar-searching pipeline.

Pulsar radio emission is a broadband signal originating from
kpc distances.
Therefore, the signal is dispersed by the ionized interstellar
medium. This causes the low-frequency components of a signal to arrive later than 
the ones at higher frequency. The delay between the two frequencies $\nu_1$
and $\nu_2$ is proportional to 
DM$(\nu_1^{-2}-\nu_2^{-2})$ where DM, the dispersion measure, is the column
density of free electrons along the line of sight. This is non-zero and 
remains nearly constant for a pulsar. 
Because of this, the wide-band signals of pulsars
need to be recorded into narrow frequency channels to prevent smearing by
dispersion.
{\it PRESTO} first searches for narrow band or non-dispersed
periodic signals in the raw channelized data and removes them, since they are likely terrestrial. Then it generates time series for an array of DM values
by adding appropriate time delays to each frequency channel.
The range of DM searched is $0<$DM$<5000\,{\rm pc\cdot cm^{-3}}$, which easily
encompasses
the expected Galactic interstellar dispersion for all lines of sight in the PALFA survey
\citep{cl01} and also maintains sensitivity to any possible highly-dispersed
extragalactic radio sources (e.g. \citealt{lbm+07,tsb+13}, L. Spitler et~al. in prep.). 
After that it searches for periodic signals in the ``de-dispersed''
time series using a Fourier Transform and harmonic summing, and picks out the significant periodicity
peaks in the power spectrum (see
\citealt{ran01} for details). For each candidate periodicity, {\it PRESTO} folds the
de-dispersed time series into a 3D data cube (time interval, phase and channel frequency) 
using the period, and stores the folded data in a {\it pfd} format
file together with data descriptors such as the date and coordinates
of the observation. These {\it pfd} files can
later be converted to candidate diagnostic plots using routines in the {\it
PRESTO} software suite.

Direct inspection of the folded 3D data array is inconvenient, so it is usually
projected into several lower-dimensional plots. The {\it PRESTO} routine
{\it show\_pfd} is designed to display the {\it pfd} file as a candidate plot that
contains several of these projections. Figure~\ref{fig:presto} is an example
pulsar candidate plot, with the most important subplots highlighted.

1.  {\bf Summed profile:} One can sum all frequency channels and time intervals to
create a summed intensity-versus-phase pulse profile. Pulse profiles of real
pulsars are usually composed of one or several very narrow peaks, though there
are some known pulsars with pulse profiles which are broad and/or contain multiple
peaks.

2.  {\bf Time-versus-phase plot:} This plot is obtained by summing the
data over the different frequency channels, leaving the pulse profile as
averaged over subintervals of the observation. One or more vertical stripes in this plot
indicates that a pulsed signal was observed for the duration of the scan. 

3.  {\bf Frequency-versus-phase plot:} Summing the data cube over the
different time intervals leaves the frequency-versus-phase plot. The
presence of one or more persistent vertical lines in this subplot, as in the
example, indicates a broadband signal during the pulsed emission, as
expected for a pulsar candidate. However, scintillation caused by the
interstellar medium may affect a pulsar's signal, degrading the signal
in some frequency channels.

4.  {\bf DM curve:} The plotting program searches over a range of DMs around
the best reported value. For each DM trial, it de-disperses the data cube
accordingly and calculates the $\chi^2$ of the de-dispersed pulse profile
against a horizontal line fit. The DM curve is a plot of the trial DMs
against their corresponding $\chi^2$ values. A large $\chi^2 $ value indicates
that the periodic signal deviates strongly from simple white noise. The DM curve of a real
pulsar will likely peak at a non-zero value unless affected by strong RFI.

These are the four most important features that human experts look at when
classifying candidates, and they are the inputs to our AI system. 

\subsection{Data preparation    }
\label{sec:data}
The application of pattern recognition to pulsar candidates is not a trivial
task. This is because the integration time of survey
observations may vary, and the number of phase bins with which the data are
folded also changes depending on the period of the candidate. As a result, 
the number of data points in the subplots can vary from candidate to
candidate. For ML to work, we need to
carefully prepare these data, and make sure the input data blocks for a particular
classifier have identical shape and size.

We extract the key feature plots from the {\it pfd} files. These plots consist
of 1D data arrays (summed profile or DM curve) and 2D data (time/frequency-versus-phase plot). The sizes of the data arrays vary from candidate to
candidate. For example, some candidates have 64 phase bins, while others may
have only 32 bins; some candidates have 50 time steps, while others have 100. 
This is a result of the search pipeline, which uses fewer phase bins
and more time steps for short period pulsars, and more phase bins with
fewer time steps for long period pulsars---in either case, the number
of pulses coherently added in each time interval is roughly the same.
For ML to work on these plots, the features should have the
same size and scale, so we down-sample or interpolate the data
to a uniform size: 64 bins for the summed profile, 64$\times$64 (or
48$\times$48 depending on the
classification algorithm, see Table~\ref{tab:clfs} for details) bins for
the time/frequency-versus-phase plot, and 60 bins for the DM curve. 
Piecewise linear interpolation was used for the 1D data and spline
interpolation\footnote{The 2D interpolation routine in
scipy.ndimage.interpolation.map\_coordinates was used.} for the 2D data.
We also
normalize the data to zero median and unit variance to remove the absolute scale
of the plots. For the 2D image arrays, normalization is performed line-by-line
along the phase axis, which removes instrumental variations over the
course of the observation and across the observing band, but maintains
the variance in signal across the phase---this should be dominated by
the pulsar signal.

\begin{figure}
\includegraphics[scale=0.6]{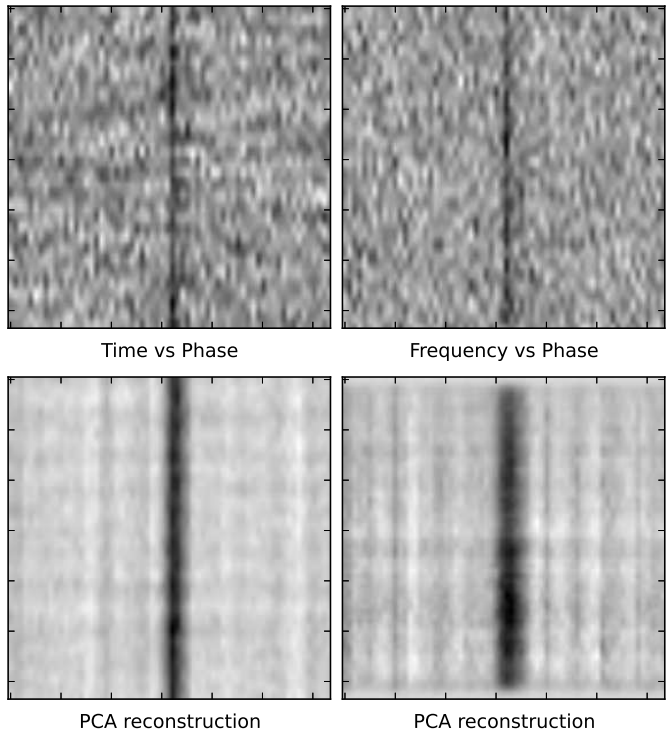} \\ 
\caption {\label{fig:pca} 
Top: The original time-versus-phase and frequency-versus-phase plots from a pulsar
candidate.
Bottom: The PCA-reconstructed plots from the top 24 PCA components.
}
\end{figure}

Once the plots are resized and normalized, we can use them to train a
ML system. 
However, the thousands of pixels in the 2D plots may slow down the training
process for the support vector machine (Section \ref{sec:layers}) classifiers.
In this cases the number of internal parameters requiring training is
proportional to the size of the input data, so to speed up the
computation, the image features (time/frequency versus
phase plot) are characterized by a principal component analysis\footnote{\url{http://scikit-learn.org/stable/modules/generated/sklearn.decomposition.PCA.html}}
 (PCA).
The PCA algorithm uses a singular value decomposition to compute a limited set of
basis images from the training data. An unknown candidate's data, then,
has a new representation as the set of eigenvalues resulting from the
projection onto these basis images. In this way, we can compress
the image features from thousands of numbers per candidate to only 24
numbers\footnote{The number of singular vectors used in the
  decomposition is a free parameter in the AI system, but is fixed by optimizing the performance in cross-validation tests.},
greatly reducing the number of parameters required in a
fit. Figure~\ref{fig:pca} shows the original 2D images of a pulsar
candidate and their reconstructions from the most significant 24 PCA
components. One can see that the PCA reconstructions capture the
important features in the images, especially the vertical stripes. 
By filtering out the weaker PCA components we also reduce the noise level in
the reconstructed image.
Some image classifiers trained much faster
with PCA-compressed features and perform as well as those trained with
full-sized images.
However, we did not apply the PCA-compression for the deep-neural-net
classifiers because they perform significantly better without it.

One last challenge was to prevent the AI from developing
any phase-related bias. In principle, a candidate's pulse may appear at any
phase, but in practice many candidates peak at phase 0 or 0.5. We found that
an earlier version of the AI failed to detect some good candidates
that peaked away from phase 0 or phase 0.5. We resolved this problem by 
shifting
candidates' strongest peaks to phase 0.5 before feeding them to the AI. The resulting 
AI was tested with candidates of random phase and showed no sign of
bias.

In summary, the data-preparation procedure for all candidates fed into
the AI system involves rescaling the data to zero median and unit
variance, shifting the peak intensity to a phase of 0.5, down-sampling or
interpolating onto a standardized grid and, optionally, applying PCA.

\subsection{Two-Layer AI: a committee of experts}
\label{sec:layers}

\begin{figure}
\includegraphics[scale=0.8]{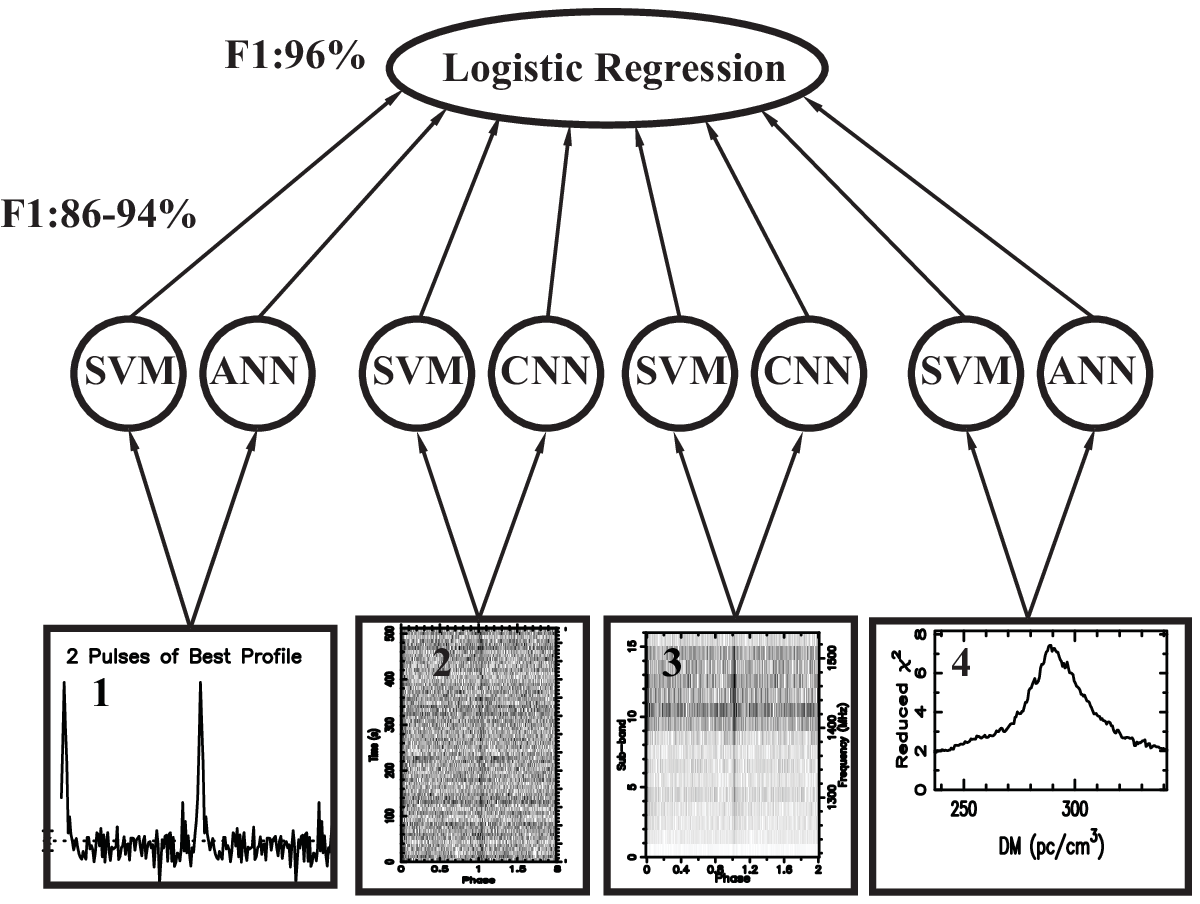} \\ 
\caption {\label{fig:layers} 
The first layer of classifiers learns how to rate each
feature to $\sim $90\%~$F_1$, and the second layer learns how to classify candidates based on the
output of the first layer.
(SVM: support vector machine, NN: Neural Network, LR: logistic regression and
adaboost are machine learning algorithms.
See Section \ref{sec:layers} for the definition of $F_1$.
)
}
\end{figure}

The first layer of \acro (see Figure \ref{fig:layers}) uses a combination of two ML algorithms on each of the four subplots, giving eight ratings in
total. The set of ML algorithms includes
artificial neural networks (ANN; see Section \ref{sec:ann} for detail), convolutional neural networks
(CNN\footnote{\url{https://github.com/aberndsen/NeuralNetwork}};
see Section~\ref{sec:cnn} for detail) and support vector machine
(SVM\footnote{\url{http://scikit-learn.org/stable/modules/svm.html}};
an algorithm that finds a direction in the parameter space on which the
distance between the two classes of data points are maximized; see
\citet{cc01a} and \citet{sklearn} for the implementation details). The choice of which algorithms to use was
determined both by their individual performance and their benefit to the
overall performance. It should be noted that the combinations were the
result of extensive testing, including using other standard ML
algorithms such as decision trees. In the end, \acro uses an ANN with
one hidden layer of logistic units and a radial kernel SVM classifier
on each of the 1D subplots (pulse profile, DM curve), and
it uses a CNN and SVM on each of the 2D subplots (pulse
interval versus phase, pulse frequency versus phase). 

In the second layer, we combine the scores from the eight first-layer
classifiers using another ML classifier (Figure~\ref{fig:layers}). 
Several algorithms are appropriate for this purpose, and we tested 
logistic regression (LR), ANN and  SVM algorithms. 
The best performance was from a
simple LR with L2 penalty\footnote{A L2 penalty means the algorithm also
tries to minimize $(\sum_i w_i^2)/C$, preventing the weights $w_i$ from
growing too large. $C$ is a small control parameter.}, and this option was chosen for the second
layer of the \acro system. The LR algorithm assigns each of the 
first-layer scores ($x_i$) a weight ($w_i$), computes their weighted sum,
and converts this sum to a probability using the logistic function 
\begin{equation}
P =\frac{1}{1+e^{-\sum_i w_ix_i}}\,.
\end{equation}
The LR algorithm finds the best set of $w_i$ that minimize the classification
errors in the training data. 

In order to test how well the classifier can perform on new data, we
employ a validation test. To do this, we first split
the labelled PALFA data into a training set and a test, or cross-validation,
set. We use the training data to train the classifiers and the validation data to
test them. Because the validation data were set aside from the beginning, this
test result is a reliable estimate of the classifier's performance. 

A commonly used performance metric is the $F_1$ score, defined as
the harmonic average of precision $p$ (the ratio of true pulsars to
the total number of candidates ranked as a pulsar) and completeness
$c$ (the ratio of ranked true pulsars to the total number of pulsars in the
validation set)
\begin{equation}
F_1 = \frac{2p c}{p+c}\,.
\end{equation}
The $F_1$ score from validation tests may vary slightly from test to test due
to random fluctuations. We run many iterations of training and testing in
order to find a reliable estimate of it. Every iteration starts with randomly
splitting the labeled data into new groups of training and test data, so the new
test will be different from the previous one (See Table \ref{tab:clfs}).

The classifiers in both layers of the system all have internal design
parameters that need to be fixed. This is accomplished by a grid search over the
possible values of the design parameters, and is fixed by the set 
that maximize the $F_1$ score in a validation test.
The optimized parameters of the eight
classifiers and their test performance on each feature plot is listed
in Table~\ref{tab:clfs}. The performance is also depicted in
Figure~\ref{fig:layers}, which shows the overall structure of
the \acro system.

Finally, in this particular searching-for-pulsars problem, it might
seem that we
should prefer a classifier with higher completeness than precision. 
In general, maximizing the precision alone
would result in a cautious AI system which would miss a lot of true
pulsars, while maximizing completeness would bring
in a lot of false positives. When training the PICS, we preferred 
a balanced AI that maximizes $F_1$.
This is because we want
to improve the AI's completeness when we add more varieties of
pulsars to the training data and we also want to improve its precision when we
add more RFI candidates, to emphasize on one metric will hinder our ability to
improve the other.
When applying the AI in practice, we can adjust its completeness and
precision to our needs by changing the cut on $P$, the AI probability score.


\subsection{The neural networks}
\label{sec:ann}
In the first layer of PICS, we used the ANN for the 1D subplots and the CNN for
the 2D subplots, they are different types of neural network.  In this Section, we briefly introduce some
terminology common to the understanding of both these neural networks.
For the detail implementation of LR and SVM please refer to \citet{sklearn} and \citet{cc01a}.

Biologically, a neural network is a collection of neurons connected by
synapses, where individual neurons respond, or ``fire'', to different
inputs. An artificial neural network is the computer analog, modelled as a 
function of many inputs (synapses) to produce a single output. These
functions are often called activation functions $h$, and typically
have finite range for classification purposes. The most
commonly used activation functions are the logistic (sigmoid) function
$h(x) = 1/(1+\exp(-x))$ and the hyperbolic tangent function $h(x) =
\tanh(x)$. The former maps any float input to a number between 0 and
1, the latter maps to a number between $-$1 and 1. 

In the ANN, these neurons are distributed among different
layers. All neurons in a single layer $l$ receive the same inputs
from the previous layer ${\textbf a}^{(l-1)}$, but weights the
signals with a unique set of parameters (${\textbf w_i^{(l)}}$ for the $i^{th}$
neuron in the $l^{th}$ layer).  Thus neurons in the same layer
``fire'' under different circumstances as determined by their weights
${\textbf w_{i}^{(l)}}$
 and bias $b_i$, 
$h({\textbf w_{i}^{(l)}}\cdot{\textbf a^{(l-1)}} + b_i)$. The weights
 ${\textbf w^{(l)}}$ connect 
the outputs of a previous layer in the ANN to the input of the next, serving
the role of the synapses in the neural network. A given layer with
$N_{(l)}$ neurons accepting $N_{(l-1)}$ inputs has $(N_{(l)}+1)\times N_{(n-l)}$
weights, or synapses, which makes the number of synapses in a given 
network much larger than the number of neurons. A synapse, however,
controls how much weight is given to a single input, so the collection
of synapses forms the pattern which will produce an activation. Since
there are so many synapses in an ANN, there are a lot of patterns that
the system can respond to. In practice the patterns are not known, so
the weights are initialized randomly and are determined through the training
process.

We train the neural net with manually labeled candidates through
back-propagation. The goal of training is to minimize the difference
between the neural net output and 
the human classifications $y$ (0 for RFI and 1 for pulsar) of the candidates. This difference, the prediction error
$\delta^{(l)}=a^{(l)}-y$, can be formed for each layer $l$, and the
correction to the $j^{th}$ weight  in the $i^{th}$ neuron $\Delta_{i,j}^{(l)}$
is determined as a function of ${\textbf a}$ and $\delta_j^{(l+1)}$. The exact
functional form depends on the choice of activation function and error function.
In practice, $\Delta_{i,j}^{(l)}$ is often averaged over an ensemble of
candidates, since training is done in batches for computational
efficiency.

\subsection{The convolutional neural network}
\label{sec:cnn}

The best individual classifier in the collection of experts is the
convolutional neural network trained on the frequency versus
phase plot. This deep, 5-layer network is similar to
LeNet-5~\citep{LeCun98}, a system proven to be very successful in
recognizing handwritten digits. This CNN also represents the state of the
art in machine learning and, as such, warrants a detailed description
of its structure and an explanation of its superior performance in the
candidate-selection process.

The CNN is a powerful pattern recognition 
system capable of analysing large 2D images directly without any compression.
Therefore, PCA compression, which was necessary for the SVM image data inputs, was
not applied to the inputs of the CNN.

\begin{figure}
\includegraphics[scale=0.6]{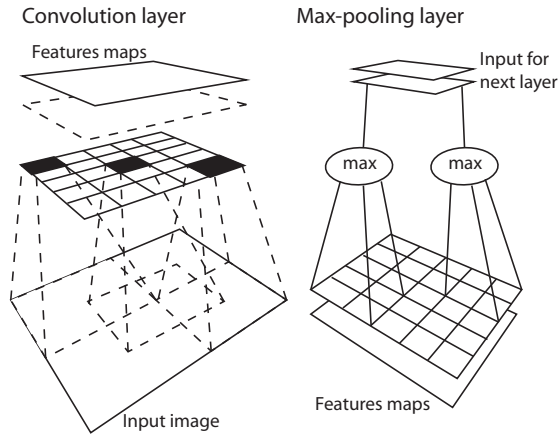} \\ 
\caption {\label{fig:cnn} 
A schematic of the first two layers in the convolutional neural
network. 
Left: convolution layer; 
from bottom up, the input image is convolved with a set of image kernels,
forming feature maps that show the presence of certain features in
different positions of the image. 
right: max-pooling layer;
the feature maps are compressed to smaller size by taking in only the maximum
values of adjacent pixels.
The PICS CNN consists of two sets of alternating convolution and max-pooling
layers, and a final artificial neural network layer.
}
\end{figure}

In the first layer of the CNN, the input image is fed to groups of neurons
with shared weights. We scan the input image with a sliding window to get 
a set of sub-images. The neurons with shared weights each look at one of these
sub-images and become active if certain feature as defined by the shared
weights is detected from it. The different groups of neurons are used to
detect different features. This process can also be viewed as convolving the input image
with a set of small image kernels, or features, and then applying the hyperbolic
tangent activation function to form a set of feature maps. Specifically, the
$k^{\rm th}$ feature map is given 
by
\begin{equation}
h^k=\tanh\left((W^k\ast x) + b\right)\,,
\end{equation}
 where $W^k$ is the
feature kernel, $x$ is the input image, and $b$ is a bias term. The
number of kernels, $k$, is a free parameter in the system, and it
determines the richness in the representation of the data. The second layer
of the CNN is a {\it max-pooling} process. This is a form of down-sampling
which bins the 2D feature maps and chooses the maximum value within
each bin. The main advantage is a reduction in computational
complexity in subsequent layers, though pooling also has the advantage
of introducing translation invariance of the feature $W^k$ across the
bin~\citep{boureau2010theoretical}.  These two steps are illustrated in
Figure~\ref{fig:cnn}. 

The output of the max-pooling layer is fed into another convolution
layer for feature detection and yet another max-pooling layer. This 
output is fed into the fifth and final layer, a traditional,
fully-connected ANN. While the previous four layers function to
locate small-scale features across the input images, the final layer
combines this information to detect large-scale features and develop
a global understanding of the original image.

The CNN configurations for both the time-versus-phase plot and
frequency-versus-phase plot were the same. In the first step
(Figure~\ref{fig:cnn} left panel), both CNNs take input images down-sampled
to 48$\times$48 pixels, convolve them with 20 different 16$\times$16
image kernels, producing 20 feature maps of size 33$\times$33. The
subsequent max-pooling layer (Figure~\ref{fig:cnn} right panel)
divides each map into 3$\times$3 boxes and compresses the convolved
image to a size 11$\times$11 by taking the maximum in each box. The
second convolution layer convolves 50 $8\times$8 image kernels
shared across the 20 feature maps, resulting in 50 different 4$\times$4
feature maps.  A second 2$\times$2 max-pooling layer further
compresses each feature map to an images of size 2$\times$2.  The result
is an array of 50$\times$4 numbers characterizing the local,
kernel-sized features of the
original image. The final layer is a traditional, fully-connected
artificial neural network consisting of 500 hidden logistic
units which take these 200 numbers to compute one final score. 

There are 8820 artificial neurons distributed through the five layers of the
CNN including the ones in the image kernels. 
All neurons in this network are hyperbolic tangent functions except the one
that forms the final output, which is chosen to be a sigmoid function in order
to map onto the classification labels 0 and 1. 
The training of the neural net is to let these synapses learn and 
store the patterns distinguishing pulsars from RFI. As an
example, the last layer has $10^5$
weights connecting the 200 outputs of the second last layer to the
500 hidden neurons in the last layer. Each hidden neuron takes, as
input, a weighted sum of the 200 outputs of the previous layer.  All
of the connection parameters are initialized randomly, 
and then updated through the process of back-propagation.

The structure of the CNN is determined by the choice of image size,
kernel size, number of kernels, pooling size, and neural network size,
these parameters are often called neural net design or hyper-parameters. 
The best CNN design, as described above, is determined through
cross-validation tests.
The labelled data are randomly shuffled and split into a training set formed from 60\% of the
candidates and a validation set from the remaining 40\%, and a large
grid search is performed. For each choice of design parameters the CNN
is trained and the performance is characterized by computing its $F_1$
score (see Section~\ref{sec:layers}) on the validation set. 
The CNN we used gives $F_1=92$\% when trained and tested on
the time-versus-phase plots, and $F_1=94$\% for the
frequency-versus-phase plots (Table~\ref{tab:clfs}). After optimizing
the CNN design, the final AI is trained with all training candidates
in order to maximize its sensitivity. The final performance of the PICS
system is tested with a set of completely independent GBNCC data (see
Section~\ref{sec:res}). 

This CNN performs well on the frequency-versus-phase plot
because a large fraction of RFI is narrow-band emission. For
continuous emission this RFI shows up as a horizontal line, while
burst-like or periodic RFI appears as a small dot. This is opposed to
the broadband pulsed signal of a pulsar, which shows up as a vertical
stripe. Since the CNN excels at detecting small-scale features, it can
easily detect this form of RFI. A second reason this classifier is the best
discriminant of pulsars is simply because there is more information in
the 2D plots than the 1D plots. 
For similar reasons, the algorithm can recognize and reject burst-like RFI or signals that drift around in phase using the time-versus-phase plot. 

This deep neural net is implemented using {\it Theano}~\citep{theano},
a computation library for python which compiles numerical expressions
to run efficiently on either CPU or GPU architectures. 

\section{Results}
\label{sec:res}

We trained the \acro AI system with 3756 labelled PALFA candidates, among them
1659 are pulsars and their harmonics and 2097 are non-pulsars. 
There are only a few thousand known pulsar/harmonic candidates in PALFA but
millions of non-pulsars. We picked similar numbers of pulsars and non-pulsars
from the PALFA candidate pool as  classified by other human experts. 
The pulsar candidates include unique pulsars and also the same pulsar with
different beam offsets. To ensure that the AI is sensitive to both pulsars and
their harmonics, $\sim40$\% of the training pulsar candidates
are harmonics of known pulsars. 

We also manually rated 
how ``pulsar-like'' is each of the four subplots for every training candidate
regardless of the candidate's nature. This is because some RFI candidates may 
have a good pulse profile or DM curve, and we want to train the AI to
recognize those the good subplots even when the candidate itself is
not a pulsar. The first layer classifiers are designed to recognize 
patterns in individual subplot in the diagnostic plots of 
the candidates. Therefore they are trained to predict the ``ratings'' of the 
subplots for the training candidates, not to predict the classification of the
candidates.
In contrast, the second layer (LR) classifier is designed to predict whether
or not a candidate is a pulsar based on the output from the first layer
classifiers. Thus, it is trained with the classifications of the candidates,
not with ``ratings'' of their subplots.

To ensure that the trained classifiers can work on new data and to determine
their best initial parameters, we performed standard cross-validation
tests (Section~\ref{sec:cnn}). 
We shuffled and split the 3756 candidates into two groups, 60\% for training and
40\% for validating. Furthermore, we repeated this shuffle, split, train and validate
procedure 10 times to make sure their performance on validation data is
reliable and repeatable. The individual first-layer classifiers score
in the range 86--94\%~$F_1$ (with less than 1\% RMS each) on the
validation data (Table~\ref{tab:clfs}). The second layer of PICS, a
LR algorithm that combines the scores from first-layer classifiers to
form a consensus vote, scored an average of 96\%~$F_1$ (8\% misses, 1\%
false positives) on the  validation data.  

To further measure the AI's performance and determine whether it can
be generalized to other surveys, we
applied it to a large set of GBNCC candidates that was never seen by the AI
system during training.
Like PALFA, the GBNCC survey uses the {\it PRESTO} search pipeline and generates
candidates in {\it pfd} format, but this survey was conducted using the
Robert C. Byrd Green Bank Telescope (GBT) instead
of Arecibo, so the RFI environments of the two surveys are expected to be
different; furthermore, GBNCC survey uses the 350~MHz receiver instead of the
1400~MHz one, thus the 
DM curves of the GBNCC candidates are expected to be slightly sharper than that 
of the PALFA candidates thanks to the more significant dispersion effect in
the lower frequency band.
We applied the AI to 90008 manually labelled GBNCC candidates.
An initial test showed that the AI can sort all 56 pulsars in this data set to
the top 3.8\%, and 68\% of the pulsars to the top 0.16\%. The first
RFI appeared in rank 136, following 29 pulsars and 106 harmonics.

\begin{figure}
\includegraphics[scale=0.4, angle=270]{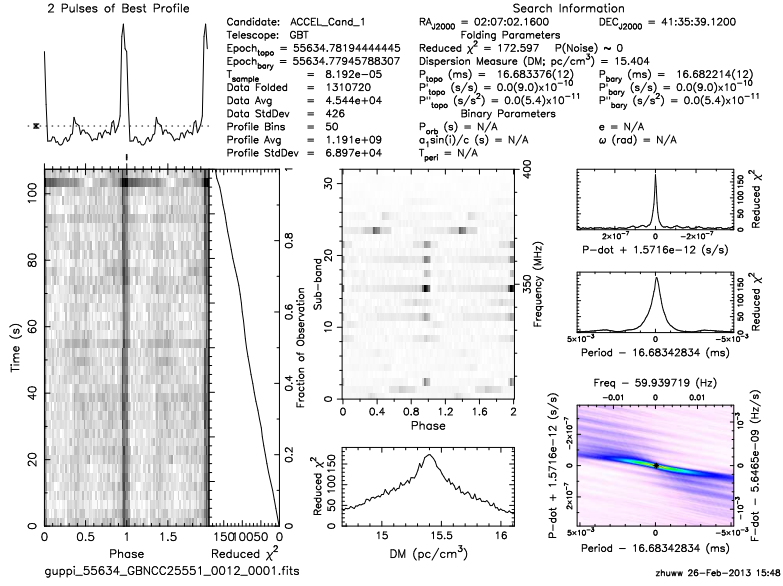} \\ 
\caption {\label{fig:rfi} 
An example of a $\sim$60~Hz RFI signal that was ranked higher than the weakest four  
pulsars in the GBNCC test data.
}
\end{figure}

An inspection of the false positive candidates revealed
many to be harmonics of the 60~Hz power signal (see
Figure~\ref{fig:rfi} for an example). Table~\ref{tab:rfi} lists the 6
most-populous frequency ranges in a frequency histogram of all candidates, the
majority of these candidates are likely caused by the 60~Hz power supply.
Although these frequencies comprise $<$1\% of the frequency domain searched by
{\it PRESTO}, 24519 (roughly 27\%) of the candidates fall into these frequency
ranges. 
It is standard practice to remove harmonics of 60~Hz for North
American surveys; in spite of these efforts, a significant amount of
such RFI remains. Using the image patterns in the diagnostic plots,
our AI was able to reject majority ($\sim96.2$\%) of this RFI, with
the remaining false positives strongly resembling pulsars (see Figure~\ref{fig:rfi} for an example).

In light of this observation, we adjusted the scores of all candidates using a
Bayesian prior on the pulse frequency $f$ to reduce the score for the
false-positive candidates in contaminated frequency bins:
\begin{equation}
P(p|f) = \frac{P(p)}{P(p)
  + \left[P(f|r)/P(f|p)\right]P(r)}\,.
\label{eq:prior}
\end{equation}
Here $P(p)$ is the probability of being a pulsar as previously scored
by PICS, $P(r) =1-P(p)$, while $P(f|r)$ and $P(f|p)$ are the probability
density functions, or likelihoods, of RFI and pulsars in frequency. 
$P(f|r)$ can be
well-approximated by sampling the frequency distribution of all test 
candidates, which is dominated by non-pulsar signals, especially those at 60~Hz and its harmonics (Table~\ref{tab:rfi}).
We approximate the
prior distribution $P(f|p)$ by median-filtering $P(f|r)$, which
removes spikes caused by RFI and leaves a distribution that reflects the survey sensitivity. 
At the
harmonics of 60~Hz $P(f|r)\gg P(f|p)$, 
such that $P(p|f) < P(p)$, and the AI score is reduced. 
Away from the few affected frequency bins (Table~\ref{tab:rfi}),
$P(f|r)\simeq P(f|p)$ and the AI score is lightly affected, with
$P(p|f)\simeq P(p)$.  

The frequencies of the test candidates are binned from 0~Hz to 2000~Hz in steps of 0.5~Hz. Due
to the limited bin size, all slow pulsars ($f<$1~Hz) are
binned to the lowest two bins, along with a lot of RFI. Since the underlying structure of the
prior distributions is not properly resolved in this domain, we did not apply the Bayesian rules
to candidates with $f<$1~Hz.

Figure~\ref{fig:GBNCC} shows the distribution of GBNCC test candidates when
sorted by $P(p|f)$, and summarizes the performance of PICS. After
applying the Bayesian rule (Equation~(\ref{eq:prior})), 
all 56 pulsars and 208 of the 221 harmonics were sorted and placed into the
top 961 (1\%) of the 90008 candidates by the PICS system. 
The first RFI candidate appeared rank 187, following 45 pulsars and 141
harmonics.
The 13 harmonic candidates that were not sifted to the top are mostly
candidates with very low SNR and broad profile.

A blind cross-validation test was performed using pulsars and RFI candidates
collected from the PALFA survey but never seen by the AI.
We found that the same Bayesian rule (Equation~(\ref{eq:prior})) that worked
for the GBNCC test also clearly reduced the false positive rate in this
cross-validation test.
This indicates that the Bayesian rule can be generalized to surveys other than
GBNCC. 
Caution must be advised, though, since a weak pulsar candidate falling in an
RFI-contaminated frequency bin (Table~\ref{tab:rfi}) will be down-weighted by the
Bayesian prior and could be rejected. Fortunately, the chance of this happening is
very small, since only
a few frequency bins ($\sim 0.7$\%, Table~\ref{tab:rfi}) are affected significantly by the prior. In light of this, the Bayesian-rule
RFI rejection is included only as an option in PICS and can be turned off.


Despite the size and complexity of PICS, candidate classification is
a fast process because most of the individual
classifiers simply apply dot products. The hard part of the computation
was already done during the training phase. 
It took $\sim45$ minutes to classify the 90008 GBNCC candidates using
a cluster of 24 2.7GHz CPUs, $\sim$ 0.7 CPU second per candidate, though most of
the time was spent on disk I/O and not all CPUs were used at 100\% capacity. 
 Using the same computer
cluster, the AI would be able to classify a million candidates in several
hours.

\begin{figure}
\includegraphics[scale=0.4, angle=270]{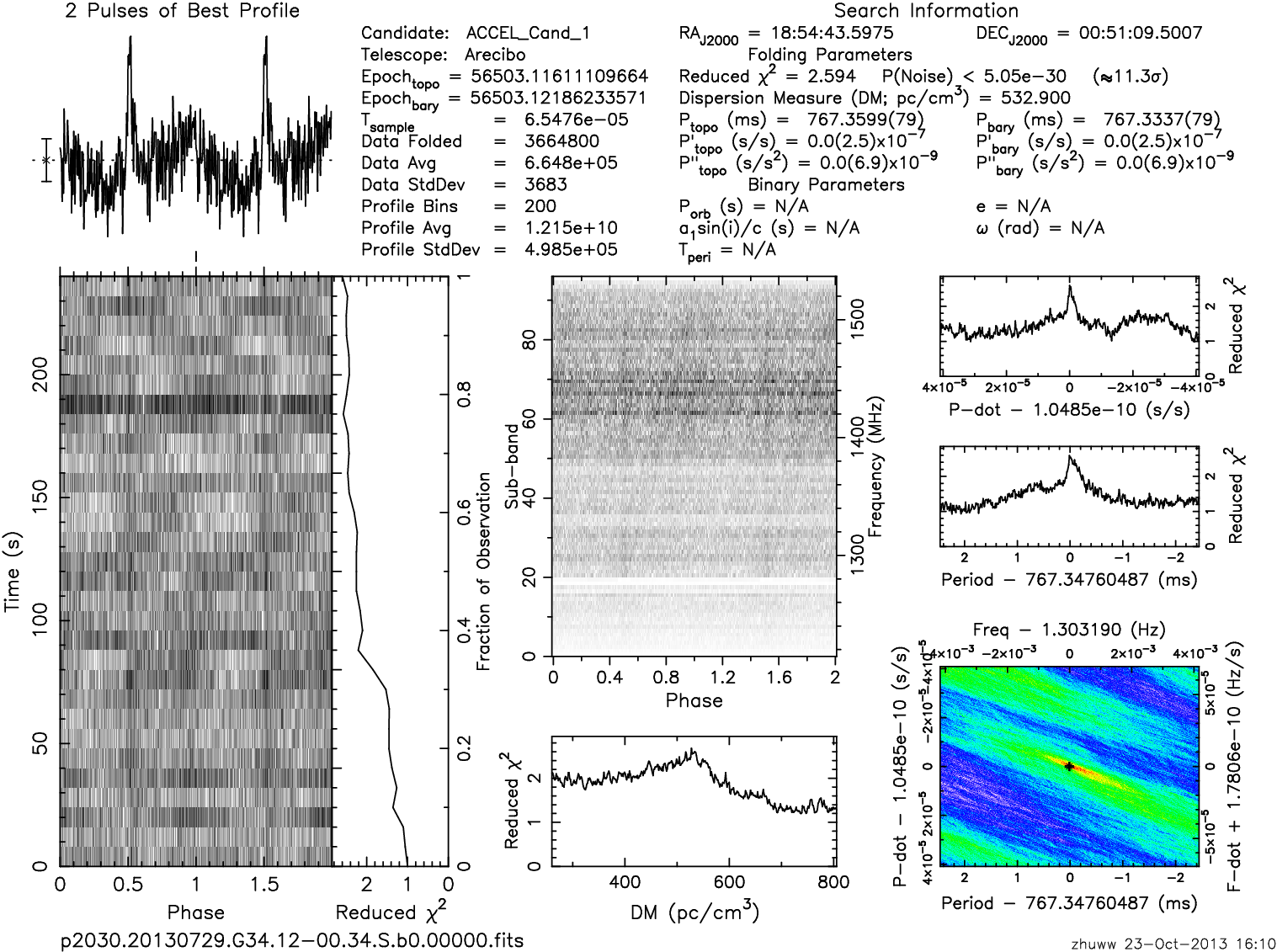} \\ 
\caption {\label{fig:1854} 
The discovery  plot of PSR J1854+00.
}
\end{figure}

The \acro AI has been integrated into the PALFA pipeline, such that we can
query and sort candidates using the AI rating on the {\it cyberska.org} web
platform. With the help of the AI, we have found many promising candidates
over
several weeks, six of which have been confirmed as new PALFA discoveries. 
Here we list the discovery parameters of these pulsars:
\begin{itemize}
\item PSR J1914+08 (Figure~\ref{fig:presto}) is a 146.68~ms pulsar with a DM of
289~pc/cm$^3$.
\item PSR J1938+20 (Figure~\ref{fig:1938}) is a 2.634~ms MSP with a DM of 237~pc/cm$^3$.
\item PSR J1901+02 is a 885.24~ms pulsar with
a DM of 403~pc/cm$^3$.
\item PSR J1903+04 is a 1151.39~ms pulsar with a DM of
473~pc/cm$^3$
\item PSR J1930+17 (Figure~\ref{fig:1930}) is a 1609.72~ms pulsar with a DM of
232~pc/cm$^3$
\item PSR J1854+00 (Figure~\ref{fig:1854}) is a ~767.334s pulsar with a DM of
533~pc/cm$^3$
\end{itemize}
These pulsars are now the subject of on-going timing observations by Arecibo or
Jodrell Bank Observatories.

\section{Discussion}
\label{sec:dis}
\begin{figure}
\includegraphics[scale=0.6]{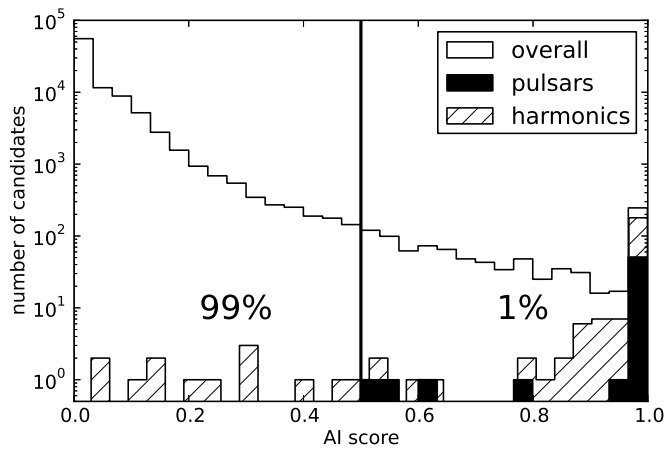} \\ 
\caption {\label{fig:GBNCC} 
The unfilled histogram is the distribution of AI scores of 90008 GBNCC
candidates. The filled histogram is the AI score distribution of known pulsars. 
The hatched histogram is that of the harmonics of known pulsars.
}
\end{figure}

\acro scores the
candidates with a number between 0 and 1, a higher score corresponding
to a more pulsar-like candidate.
When tested with the GBNCC data, the AI 
placed 100\% of the pulsars and 94\% of the harmonics in the top 1\% (Figure
\ref{fig:GBNCC}) of ranked candidates. 
By rejecting 99\% of the
candidates, PICS can improve the speed of the
candidate classification process for a human expert by a factor of $\gtrsim$100.
If we combine the AI with
other ratings or scores such as the signal-to-noise
ratio, the sorting efficiency can be further improved, making it possible
for a few human experts to sift through millions of candidates. Such a system
can be very useful for existing and future pulsar surveys.

In the GBNCC test, PICS AI showed very good performance.
However, in the top 1\% of the test candidates, there were still a few pulsars 
ranked below hundreds of non-pulsars. These were pulsars with broad
pulses and low signal-to-noise ratios, and were surpassed by some
strong RFI signals that also have broad features. 
It seems that the PICS AI could benefit from training with more pulsar/RFI training candidates that have broad pulse profiles.

Despite being limited by the quantity and diversity of the training
data, the use of image-pattern-based machine learning in PICS is a novel idea
and has some advantages:

The AI uses an ensemble of ML algorithms, including a deep neural network composed of many neurons and hidden weights which have the capability to recognize subtle or complex features.
By gathering more human-identified candidates from surveys, we will be able to
further improve the AI system. 
Specifically, if a survey encounters a new kind of RFI that our system or
a method based on analytical heuristics
fails to reject, we can improve our AI by incorporating
examples of these RFI into our training data and retrain the system.
Future improvements to \acro
include expanding the training set to candidates 
from both PALFA and GBNCC surveys, and providing input capability for survey
products from other search software. Rather broadly, by increasing the pool
of training candidates, we can improve the accuracy of the \acro AI.

The \acro AI makes a classification based solely on image patterns in the
diagnostic plots. A caveat---and, possibly, feature---is that because these plots are re-binned
and normalized to unit variance, the AI ignores the DM, period and pulse
amplitude differences between different candidates.
Although this lost information may be useful
in some rare cases (e.g. ruling out millisecond candidates with high DM based
on the DM smearing timescale), 
this procedure forces \acro to learn the universal features of pulsars
(broadband, pulsed and dispersed emission) and helps it 
adapt to other surveys for which sensitivity in candidate DM,
frequency, and signal strength may differ. \acro is also forced to rely
less on information than other score-based systems rely on, making it a good complement to them.

\begin{figure}
\includegraphics[scale=0.4, angle=270]{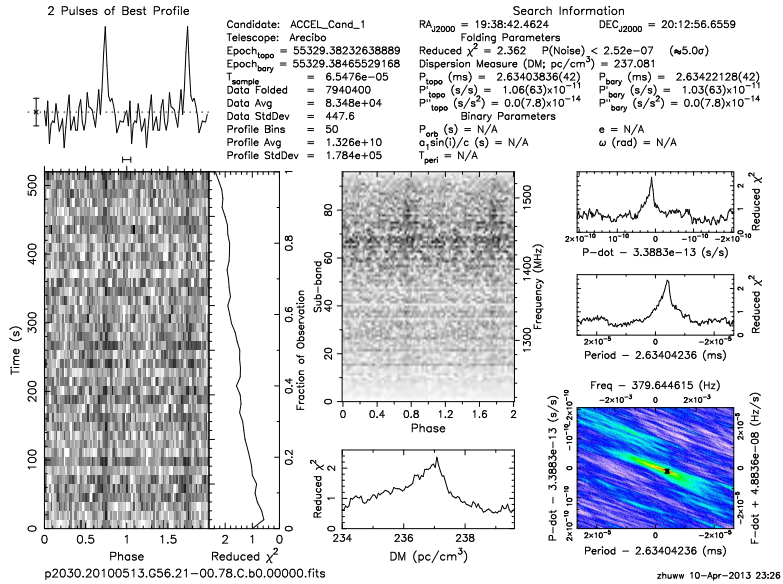} \\ 
\caption {\label{fig:1938} 
The discovery  plot of PSR J1938+20.
}
\end{figure}

Being trained with many harmonic candidates with weak or multi-peak
pulse profiles, the
PICS AI is also very good at finding such candidates.
\acro has discovered six new pulsars since being integrated
into the PALFA survey pipeline. 
PSR J1938+20 was discovered as a very weak candidate ($5.0\sigma$
according to the version of {\it PRESTO} in the PALFA pipeline at the time of
discovery; Figure \ref{fig:1938}). This makes it the least
significant pulsar candidate confirmed by the PALFA survey. Such a candidate
probably would not have been found by a candidate-sorting method based on the significance of the signal.
PSR J1914+08 was discovered as a multi-peaked harmonic, not
as a candidate at its fundamental frequency.
Other candidate selection methods that rely on fitting the pulse profile with a
single Gaussian curve would likely down-rate these harmonics.
Due to the presence of strong RFI, PSR J1930+17's DM curve (Figure \ref{fig:1930}) is significantly skewed to the left. 
Similarly, PSR J1854+00's DM curve appears to be rather shallow and also
slightly skewed. 
It is hard to characterize these atypical DM curves using simple
pre-designed functional forms.
But our ML system was able to recognize these DM patterns based on similar
examples in the training data.

A caveat is that the current PICS AI is trained and tested with
data from PALFA and GBNCC survey, both of which have short exposure times of
several hundred seconds per pointing. The pulsar candidates observed by these 
surveys were rarely affected significantly by scintillation or binary motion.
In a survey with significantly longer exposure time, scintillation would make
the pulsar signal patchy and discontinuous in the 2D
subplots, while binary motion may cause pulsar signals in the time-versus-phase plot
to become curved, indicating acceleration.
For surveys and systems exhibiting scintillation and orbital
acceleration the AI will need to be retrained with suitable candidates. Alternatively, 
PICS could be trained from candidates generated from numerical simulation which include
these effects.

Ultimately---and with enough resources---we envision a system
that adaptively trains the AI and re-scores the survey candidates on
the fly while human experts are classifying them. Such an ``online'' system will ensure portability of the AI when the RFI environment 
of the telescope slowly changes over time. Remarkably, owing to the nature of the ML algorithms
employed, this does not require coding new heuristics to characterize
any emerging forms of RFI. 
One of challenges in implementing an online learning system is that the
training data will be very imbalanced as pulsar surveys tend to find a lot more RFI than pulsars.
\citet{lbks13} has explored the performance of some existing online 
ML algorithms in dealing with
imbalanced data streams from pulsar surveys and found promise in these methods.
Whether PICS could be adapted into a successful online learning system
will be explored in future work.

\acknowledgements
Pulsar research at UBC is supported by NSERC Discovery and Special Research
Opportunity grants and a Discovery Accelerator Supplement, by CANARIE and by
the Canada Foundation for Innovation. 
Pulsar research at McGill is supported by an NSERC Discovery grant and
Accelerator Supplement, by the Canada Research Chair program, by the
Centre de Recherche Astrophysique du Qu\'ebec, by the Canadian Insitute
for Advanced Research, and by the Lorne Trottier Chair in Astrophysics and
Cosmology.
This work was also supported by US National Science Foundation (NSF) grants 1104902, 1104617, and 1105572.
LGS gratefully acknowledges the financial support by the European
Research Council for the ERC Starting Grant BEACON under contract no.
279702.
The National Radio Astronomy Observatory is a facility of the National Science
Foundation operated under cooperative agreement by Associated Universities,
Inc.
The Arecibo Observatory is operated by SRI International under a cooperative
agreement with the National Science Foundation (AST-1100968), and in alliance
with Ana G. Méndez-Universidad Metropolitana, and the Universities Space
Research Association.

\bibliographystyle{apj}
\bibliography{myrefs,journals1,modrefs,psrrefs,crossrefs}

\begin{figure}
\includegraphics[scale=0.4, angle=270]{f7.ps} \\ 
\caption {\label{fig:1930} 
The discovery plot of PSR J1930+17.
}
\end{figure} 

\clearpage
\begin{deluxetable}{cccccc}

\tabletypesize{\footnotesize}
\tablewidth{0pt}
\tablecaption{\label{tab:clfs} Classifier parameters and test performance. }
\tablehead{ \colhead{classifier}  &\colhead{feature (size\tablenotemark{a})}
&\colhead{nodes\tablenotemark{b}}  &\colhead{$\gamma$\tablenotemark{c}}
&\colhead{$C$\tablenotemark{d}}  &\colhead{$F_1$\tablenotemark{e}}   }
\startdata
ANN&  summed profile (64)&  25&  --&  2&  0.86\\
SVM&  summed profile (64)&  --&  0.08&  1&  0.93\\
CNN&  time vs phase (48$\times$48)&  500&  --&  1&  0.92\\
SVM&  time vs phase (64$\times$64)\tablenotemark{f}&  --&  0.01&  1&  0.88\\
CNN&  frequency vs phase (48$\times$48)&  500&  --&  1&  0.94\\
SVM&  frequency vs phase (64$\times$64)\tablenotemark{f}&  --&  0.001&  24&  0.88\\
ANN&  DM curve (60)&  9&  --&  10&  0.91\\
SVM&  DM curve (60)&  --&  0.2&  25&  0.91\\
LR& layer-one pool (8)& --& --& 0.1& 0.96
\enddata
\tablenotetext{a}{The input features were down-sampled or
interpolated to a uniform size.}
\tablenotetext{b}{Number of nodes in the hidden layer of the neural network
(NN). For the CNNs, only the number of nodes of the last hidden layer is
listed. See Section \ref{sec:cnn} for more details.}
\tablenotetext{c}{$\gamma$ is the radius parameter for the {\it rbf} kernel
function used in the  SVM. 
The traditional SVM can only use linear boundaries to
classify data. The kernel functions enable the SVM to find curved
boundaries. The {\it rbf}-kernel SVM is considered a more effective classifier
than the traditional SVM. 
}
\tablenotetext{d}{The $C$ parameter controls the regulation in the ANN
  and the SVM.
Here regulation means that the ANN or SVM tries to minimize a penalty function 
while finding the optimal set of internal parameters $w_i$. 
The function is often defined as $(\sum w_i^2)/C$ in order to penalize very large weights.
}
\tablenotetext{e}{The average $F_1$ scores of the classifiers from 10
independent trials.
In each trial, we randomly shuffle our known candidates before splitting them
into training and testing data, after which training is commenced. The standard deviations of the $F_1$ scores are $<$0.01.}
\tablenotetext{f}{PCA compression was applied to prepare 2D image data 
only for the SVM classifiers. We compress the 64$\times$64 image into 24 PCA components.}

\end{deluxetable}


\clearpage
\begin{deluxetable}{ccc}

\tabletypesize{\footnotesize}
\tablewidth{0pt}
\tablecaption{\label{tab:rfi} The 60~Hz RFI and its harmonics. }
\tablehead{ \colhead{frequency range (Hz)\tablenotemark{a}}  &\colhead{Counts\tablenotemark{b}}  &\colhead{Percentage\tablenotemark{c}}   }
\startdata
18$-$19&  2101&  2.3\%\\
58$-$62&  12941&  14.4\%\\
118$-$121&  4480&  5.0\%\\
139$-$140&  2157&  2.4\%\\
179$-$181&  1744&  1.9\%\\
239$-$241&  1096&  1.2\%
\enddata
\tablenotetext{a}{The 6 most populous frequency ranges in the histogram of candidates.}
\tablenotetext{b}{Number of candidates within the given frequency range.}
\tablenotetext{c}{Percentage of candidates within the given frequency range.}

\end{deluxetable}

\clearpage

\end{document}